# Continuous-wave coherent imaging with terahertz quantum cascade lasers using electro-optic harmonic sampling

M. Ravaro,[1a)] V. Jagtap,[1] G. Santarelli,[2] C. Sirtori,[1] L. H. Li,[3] S. P. Khanna,[3b)] E. H. Linfield,[3] and S. Barbieri[1c)],

[1] *Laboratoire Matériaux et Phénomènes Quantiques (MPQ), UMR CNRS 7162, Université Paris 7, 10, rue A. Domont et L. Duquet, 75205 Paris, France.*

[2] *Laboratoire Photonique, Numérique et Nanosciences (LP2N), UMR 5298 CNRS, Université de Bordeaux 1, Institut d'Optique, 351 cours de la Libération, 33405 Talence, France*

[3] *School of Electronic and Electrical Engineering, University of Leeds, Leeds LS2 9JT, United Kingdom*

**Abstract**

We demonstrate a coherent imaging system based on a terahertz (THz)frequency quantum cascade laser (QCL) phase-locked to a near-infraredfs-laser comb. The phase locking enables coherent electro-optic sampling of the continuous-wave radiation emitted by the QCLthroughthe generation of a heterodyne beat-notesignal that carries the amplitude and phase of the laser field. We use this beat-note signal to demonstrate raster scan coherent imaging using a QCL emitting at 2.5THz. At this frequency, the detection noise floor of our system is of 3pW/Hz and the long-termphase stability is <3deg/hour, limited by the mechanical stability of the apparatus.

[a)] Presentaddress: CNR-Istituto Nazionale di Ottica, Via Carrara 1, 50019 Sesto Fiorentino, Italy

[b)] Present address: National Physical Laboratory, Dr. K. S. Krishna Marg, New Delhi-110012, India

[c)] Electronic mail: stefano.barbieri@univ-paris-diderot.fr

Significant scientific effort over recent yearshas aimed at the realizationof terahertz (THz) frequency imaging systemsbased on quantum cascade lasers (QCLs). Even though the continuous-wave output power of these devices can be as high as tens of mW[1-3], there are still several technological issues that need to be addressed with the detection in order to realize a system that is sufficiently sensitive, as well as fast and compact. For this reason, several groups have tested different detection techniquesand configurations. Incoherent imaging systems have been demonstrated using Golay cells, pyroelectric detectors, cryogenically cooled bolometers, andcommercial focal plane array microbolometric cameras [4-7]. More recently,a QCL-based imaging system was also demonstrated using an amorphous-silicon microbolometric camera that was specifically developed for operation in the THz range [8]. Coherent imaging techniques have also been reported in the literature, includinga pseudo-heterodyne technique based on mixing between longitudinal modes of a multimode QCL [9], self-mixing[10],andexploiting the heterodyne mixing between a QCL and gas laser [11]. In the latter, the THz QCL was frequency-locked to the gas laser line in order to reduce the phase instability of the emitted radiation field and allow the application of an inverse synthetic aperture radar technique. Our work here is based on a coherent imaging technique that was first demonstrated by Loffler et al. [12], who usedan harmonic of the repetition rate of a mode-locked Ti:S laser as a local oscillator, and mixed it with a quartz-stabilized Gunn oscillatoremitting at 0.6THz. In this case, both sources were free running since their intrinsic phase/frequency stability was sufficiently high. However, recently,it has been demonstrated that although THz QCLs have sub kHzquantum noise limited linewidths[13,14], up to a Fourier frequency of ~10kHz they are affected by a ~$1/f^2$noise component, giving rise to theline broadenings and frequencydriftsthat have been observed in several heterodyne experiments [15,16].As a consequence the scheme of Ref.[12] cannot be implemented with a THz QCL without active stabilization.

Over the last few years, techniques have been developed tostabilize THz QCLs to near-IR frequency combs produced by fs-mode-locked fiber lasers [17-20].Theseexploit the sampling (electro-optic or photoconductive) of the radiation field emitted by the QCL using the mode-locked pulses from the fs-laser. In the RF domain, this sampling gives rise to a series of heterodyne beatsignalsbetween the harmonics of the fs-laser repetition rate and theTHz field. The lowest frequency beat-notesignal can be fed into standard RF electronics and used to phase-lock the QCL to the fs-laser repetition rate, thus eliminating the phase jitter between the two sources. The sub-Hz beat-note linewidth obtained with this techniqueallowsthe coherent accumulation of the QCL signal over long integration times and the achievement of high signal to noise ratios [20]. In theworkpresented here, this harmonic sampling technique has been used to implement a coherent imaging setup using a single-mode 2.5THz QCL and a frequency doubled fs-laser comb as local oscillator. We have achieved imaging with a noise detection limit of 3pW/Hz, anda long-term phase stability of less than 3 °/hour. Themaximum detection bandwidthis determined only by the sampling rateand not by the detector rise time as found when usinga Golay cell or bolometer.

The QCL used in our experiments operates at 2.5THz and is based on a 2.5mm-long, 240µm-wide, ridge-waveguide Fabry-Perot cavity that was fabricated by optical-lithography and wet-etching (details on the waveguide and active region design can be found in Ref.[3]). The QCL was kept at a stabilized temperature of 20K using a continuous-flow, liquid-helium cryostat, and driven at a constant current of 1.49A with a commercial power supply. Under these operating conditions, the emission is single-mode and the output power measured with a calibrated THz power-meter (Thomas Keating Ltd.) is 2mW. For the near-IR laser comb, we usea frequency doubled, mode-locked fs-fiber laser (Menlo Systems, M-fiber) operating at $\lambda = 780$nm and emitting a train of ~100fs-long pulses at a repetition rate of ~250MHz.

The experimental apparatusis shown in Fig. 1 and is based on two identical electro-optic (EO) detection units, labeled EO1 and EO2, that are used respectively to (i) lock the QCL frequency to the fs-laser comb, and (ii) detect the QCL beam after it has been reflected by the imaged target[12]. For our experiments, the QCL and near-IR comb beams are each split into two using separate beam splitters (labeled BS), and detection is achieved in both units by collinearly focusing the beams onto a 2mm-thick, <1,-1,0> oriented ZnTe crystal, which is followed by $\lambda/4$ and $\lambda/2$ waveplates, and a polarizing beam splitter. These elements form an ultrafast, near-IR electro-optic amplitude modulator driven by the THz *ac* field (see Ref. [17] for details of the operating principle). Assuming for simplicity that the QCL is single mode, this modulator generates two sideband combs at ±2.5THz from the comb carrier centered at 780nm (385THz). Since the comb bandwidth is approximately twice the QCL frequency, the carrier overlaps with the sideband combs producing a series of heterodyne beat-notes, oscillating at frequencies $|(\nu_{QCL} - n \times f_{rep})|$, where $\nu_{QCL}$ is the emission frequency of the QCL, $f_{rep}$ is the fs-laser repetition rate (250MHz), and *n* is an integer number [17]. Therefore, the lowest frequency beating, $f_{beat}$, corresponds to n =Int($\nu_{QCL}/f_{rep}$) ~ $10^4$ (= 2.5THz/250MHz), and has a frequency $f_{beat}<f_{rep}/2$ ~ 125MHz. This beat-note is detected using shot-noise limited balanced detection, based on a pair of Si-photodiodes, and is followed by a fast amplifier with a bandwidth of approximately 200MHz. Ultimately, the detection bandwidth of the system is limited by the Nyquist criterion to half the sampling rate, i.e. 125MHz.

To lock the QCL frequency, the beat-note generated by EO1 is compared, using a mixer, to a signal at $f_{RF1}$ ~10MHz produced by an RF synthesizer (RF1). The error signal oscillating at ($f_{beat}$- $f_{RF1}$) is then used to control the QCL current through a phase-lock loop (PLL) circuit with a bandpass of ~2 MHz, and phase-lock $\nu_{QCL}$ to the ~ $10^4$ harmonic of the fs-laser repetition rate. In Fig. 2(a) we show an example of the spectrum of the phase-locked beat signal, $f_{beat}$, acquired with a spectrum analyzer with a resolution bandwidth of 10Hz, and a THz power impinging on EO1 of 250μW. This phase-locking is critical to allow image acquisition using a conventional lock-in amplifier.

The second half of the QCL beam isfocused on the imagedtargetusing an f/1, gold-coated, off-axis parabolic mirror (identical to the one used to collimate the QCL beam).Approximately half of the THz radiation reflected from the targetis finallyfocused on EO2. As found in previous experiments [13,17-19], the QCL is affected by optical feedback which, in this case, mainly arises from the fraction of the beam reflected from the target that is transmitted through the beam splitter. As shown in Ref. [13], optical feedback has a strong influence on the QCL frequency noise. In particular, in the present case, the phase of the radiation reflected into the QCL cavity, as well as the intensity,, changed markedly when the beam was raster scanned across the imaged object owing to changes in the surface profile and morphology. We found experimentally that this effect couldsuddenly drive the QCL out of lock,thus compromising the image acquisition. To limit the effect of this feedback we inserted an optical isolator, consisting ofa wire-grid polarizer (WGP) oriented parallel to the TM-polarised light from the QCL.  This wasfollowed, after the beam splitter, by a 3.1mm-thick, quartz quarter-waveplate with its fast-axis at 45° with respect to the polarizer [13]. As a result, after the quarter-wave plate,the THz light emitted by the QCL is left circularly polarised, and the polarisation is changed from left- to right-circular after reflection from the target. Going back through the quarter-waveplate,the reflected beam recovers its linear polarisation, now orthogonal to the wire grid polariser (and to the QCL polarisation), thus providing the isolation. Using a power meter we measured an isolation of ~16dB.  However, this wasnot sufficient to suppress the random un-locking of $f_{beat}$completely during image acquisition. Therefore an additional ~10dB of attenuation was added (~5dB attenuation per pass).

For the measurement of the radiation reflected by the imaged object, we used a standard lock-in amplifier (SRS model SR830), and a reference oscillator, RF1. As shown in Fig. 1, in order to bring$f_{RF1}$= 10MHz (= $f_{beat}$)within the lock-inreference frequency range (1mHz to 100kHz),the latter was down-converted to 70kHz by mixing with another synthesiser, labelled RF2, oscillating at $f_{RF2}$= 10MHz + 70kHz. The same synthesiser was mixed with $f_{beat}$ generated by EO2 to providethe lock-in input signal.It is important to note that the phase-lockingof$\nu_{QCL}$is a crucial requirement for the realization of our coherent imaging system.Indeed, whilst in principle the free-running $f_{beat}$ generated in EO1 could be used as a reference to demodulate the output of EO2, as in Ref [12], in practice this is not possible. In fact, withoutphase-locking, the coherence between $\nu_{QCL}$ and $f_{rep}$is dominated by thelow frequency noise component of $\nu_{QCL}$, proportional to $\sim 1/f^{\,2}$, produced by mechanical vibrations and thermal and/or current fluctuations in the device[13,14]. As can be derived from the frequency noise spectral density of the QCL (identical to the present one) reported in Ref. [13], this generates an $f_{beat}$ with an "instantaneous" linewidth of the order of 10kHz on a1ms timescale, and is subject to drifts of several MHz/s [15,16,21].As a consequence the "free running"$f_{beat}$ is not sufficiently stable to provide a reference signal for typical commercial DSP lock-in amplifiers including the one used here[22].On the other hand, phase-locking of $\nu_{QCL}$ensures coherence with$f_{rep}$, eliminating their mutual jitter, and thus allows use of$f_{beat}$as a reference.

Before proceeding to the image acquisition, we evaluated the sensitivity of our EO detection. Using a calibrated power meter, we measured THz powers of 250μW and 60μW impinging on the ZnTe crystals of EO1 and EO2 respectively. Given the 2mW emitted power from the THz QCL, these values are in agreement with the attenuations from the optical elements shown in Fig.1, including the attenuator used to decrease the optical feedback and, in the case of EO2, the reflection from the flat part of a 10 cent Euro coin that we used as a test target for our imaging system. In Fig. 2(b) we show the power in dBmof the phase-locked $f_{beat}$measured at the output of EO2 with a spectrum analyzer with a resolution bandwidth of 1Hz. The THz power was progressively attenuated from 60μW to 10pW by superimposing up to 14 A4 paper sheets. Down to 2nW, the power was measured with a Golay cell detector that had been previously calibrated using a THz absolute power meter, while the two points below the 300pW detection limit of the Golay cell were obtained using calibrated attenuators. The noise floor of -140dBm is determined by the shot noise of the photocurrent generated by the 15mW of near-IR power incident on the balanced detection, and corresponds to the minimum detectable THz power of ~3pW (3pW/Hz noise equivalent power). This is consistent with the spectrum of Fig.2(a) at the output of EO1 where 250µW of THz power yielded a SNR of ~70dB in a 10Hz bandwidth, which corresponds to a noise equivalent power of ~2.5pW/Hz (the SNR scales linearly with the bandwidth).

Figs. 3(a), (b) and (c) show respectively the amplitude, power and phase image in grey-colour scale of a 10cent Euro coin, obtained by raster scanning the object in the focal plane of the mirror using a motorized XY stage. The image was obtained by continuous line scanning along the Y direction (from left to right in the figure) at a speed of 2.2mm/s, with a step between each line of 100µm in the X direction (the acquisition rate along the Y direction was set to obtain the same spatial resolution of 100µm). The lock-in time constant was 30ms, which allowed a dynamic range of 60dB, as shown in Fig.3(b). The width of the vertical lines on the left side of the coin was measured with a profilometer (top graph in Fig.3) and was found to be 160µm, showing that the system resolution is diffraction limited. Fig.3(c) displays the phase image, as recorded from the lock-in amplifier. As expected, compared to the amplitude or power plots where the edges of the reliefs appear very clear in black owing to scattering, here the various reliefs are displayed in different grey colors corresponding to different heights, and hence different phases owing to changes in the optical path. By monitoring the phase with time we measured a phase stability, with a shift of < 3 °/hour, corresponding to an uncertainty of <1µm in the determination of the profile height. This phase stability is completely limited by the mechanical stability of the experimental apparatus. We verified with the profilometer that the height of most of the features is larger than $\lambda/2 = 60\mu m$; however, the phase displayed by the lock-in amplifier is limited to $\pm\pi$, which reduces the effective length over which the phase changes continuously to $\lambda/2$. This, together with the fact that the coin is not perfectly parallel to the focal plane (a ~0.3° inclination is enough to produce a phase change of $2\pi$ across the coin), partially explains why the phase image does not display clearly the different shapes. Another difficulty is related to small details, such as the stars and stripes on the top left corner, where the size is close to the resolution limit of the system, resulting in a poor phase contrast. Moreover, as shown by the height profile in Fig.3, some of the details such as the vertical lines, have a height that is close to $\lambda/2$. In contrast, the amplitude image, resulting from scattering, allows much clearer identification of these details. Fig.3(d) displays a processed phase image where the effect of the limited phase-range of the lock-in amplifier has been partially removed by adding or subtracting $2\pi$ to a fraction of the pixels of Fig.3(c).

The imaging speed in these experiments is solely limited by the speed of the acquisition software. From Fig.2(a), an integration time of $10\mu s$ would still allow a signal to noise ratio of more than 30dB, which would allow acquisition of in Figs. 3 without degradation. This could be achieved by replacement of the XY translation stage with a fast steering mirror, enabling acquisition within a few seconds.

**Acknowledgments**

We acknowledge partial financial support from the Agence Nationale de la Recherche (project HI-TEQ), the EPSRC (UK), and the European Research Council programme 'TOSCA'. We thank Pierre Gellie for taking the scan of the profile shown in the top panel of Fig. 3.

**Figures**

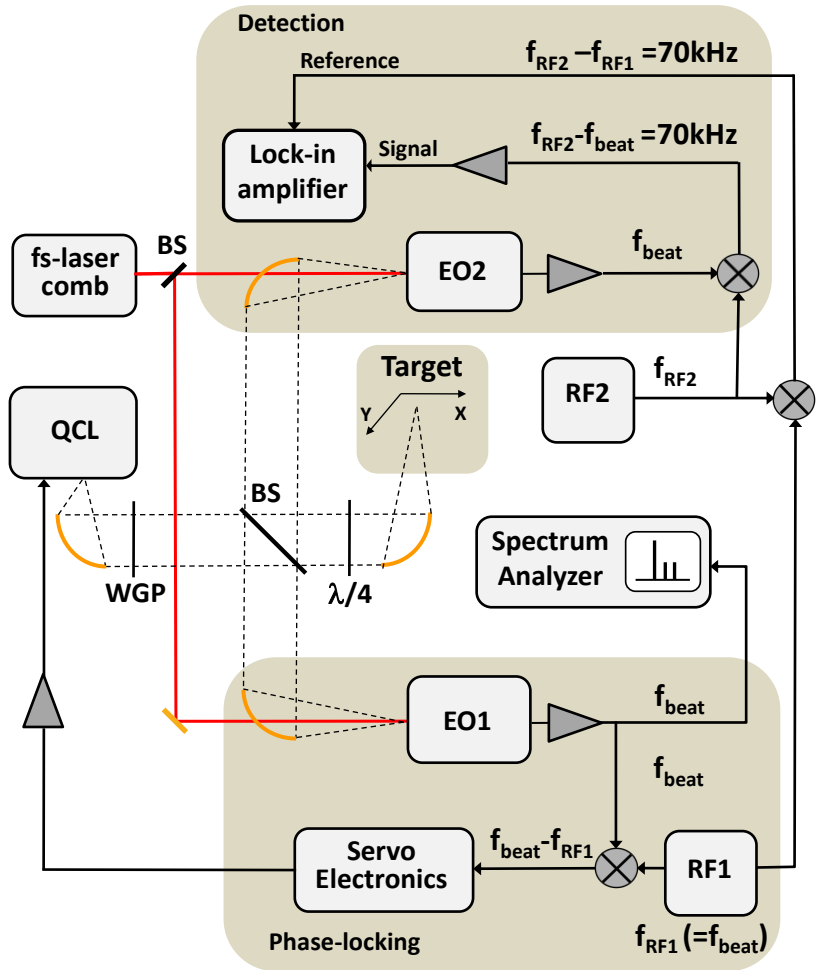

**Fig.1** Experimental apparatus. The fs-laser comb and QCL beams are split into two parts, using two beam splitters (BS), and are focused onto EO detection units EO1 and EO2. EO1 is used to phase-lock $f_{beat}$ to $f_{RF1}$, and therefore to phase-lock $\nu_{QCL}$ to ~ $10^4 \times f_{rep}$. EO2 is used to detect the THz radiation back-reflected from the imaged target. This is achieved by mixing $f_{beat}$ (= $f_{RF1}$) with $f_{RF2} = f_{beat} + 70$kHz and monitoring the difference frequency, oscillating at 70kHz, with a lock-in amplifier. The reference of the lock-in amplifier is obtained from the difference frequency $f_{RF2} - f_{RF1} = 70$kHz. All mirrors used to collimate and focus the radiation from the QCL are 90° off-axis, gold coated parabolic mirrors. The fs-laser comb is a frequency

doubled 1550nm fiber laser emitting a train of ~100fs pulses at a repetition rate $f_{rep}$ = 250MHz. There is approximately 15mW of optical power at 780nm impinging on each EO detection unit.

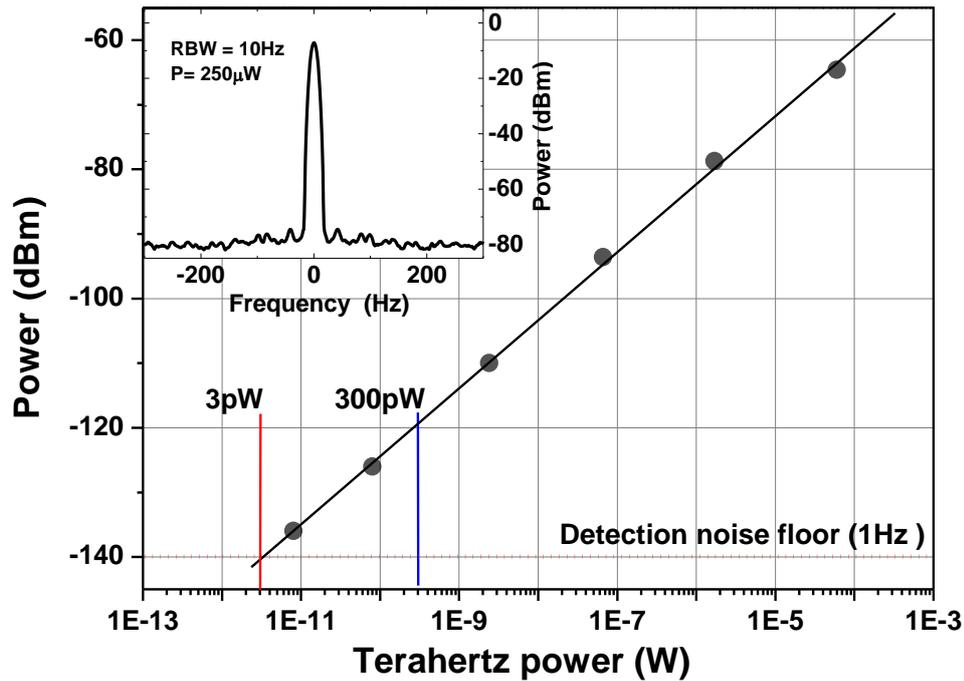

**Fig.2** Measurement of the noise equivalent power of the imaging system. The THz power focused on EO2 was decreased from 60μW to 8pW using calibrated attenuators. The noise floor at -140dBm is set by the shot noise of EO2 in a 1Hz bandwidth and gives a noise equivalent power of 3pW/Hz. Inset: Spectrum of $f_{beat}$ detected on EO1 and recorded on the spectrum analyzer with a resolution bandwidth of 10Hz. The THz power used for EO detection is 250μW.

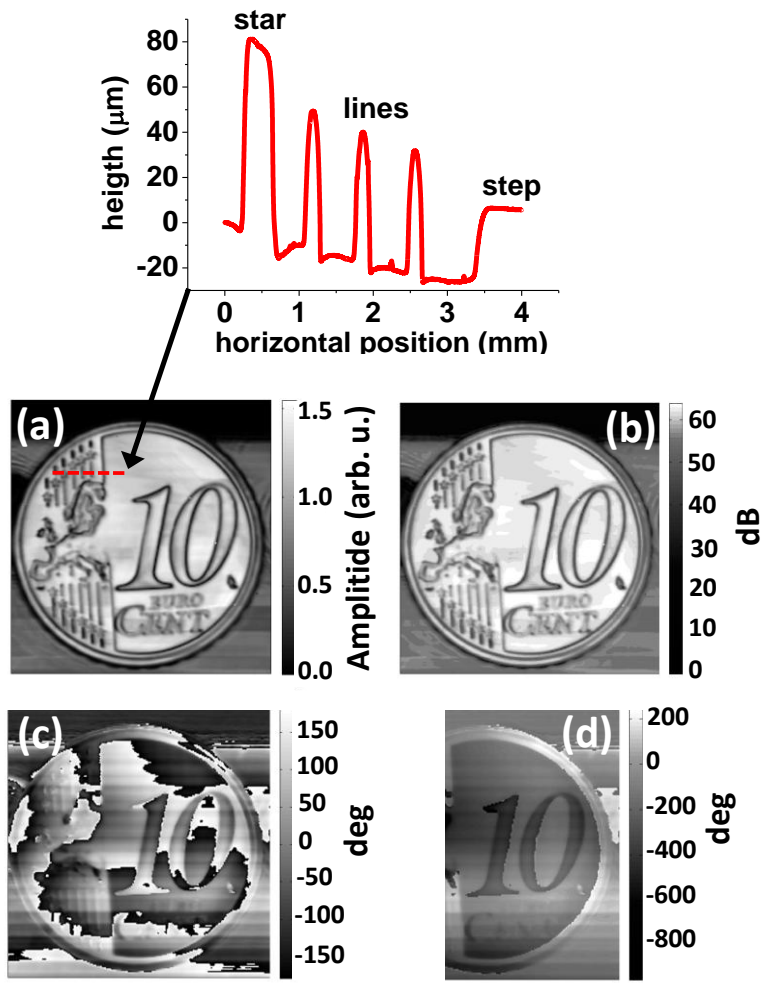

**Fig.3** THz image of a 10 cent Euro coin:(a) amplitude, (b) power. (c) raw phase image, and (d) phase image corrected for the limited phase range of the lock-in amplifier (see text). The coin diameter is 19.75mm. The imageswere acquired with lines scans from left to right at a speed of 2.2mm/s, with a step between each line of 100µmin the vertical direction. The lock-in amplifier time constant was set to 30ms. Top: Height profile obtained by scanning with a profilometer along the red dashed line of (a).

[21] Note the frequency noise presented in Ref [13] was obtained by driving the QCL with a lead-acid battery. Here the noise is significantly higher owing to the current noise of the power supply.

[22] From Ref[13], we estimate a minimum bandwidth of ~100kHz for the PLL that locks the lock-in reference to the signal reference. For the most widely used DSP lock-in amplifiers the PLL bandwidth is of the order a few tens of Hz. In addition, for our lock-in, the reference frequency range is up to 100kHz, which is well below the typical drift of $f_{beat}$, of several MHz/s.